
\magnification 1200
\parindent=0pt

\null\hskip 10cm UB-ECM-PF-95/18\par
\null\hskip 10cm October 1995\par
\null\vskip 1cm

\centerline {\bf MODE DEPENDENT FIELD RENORMALIZATION AND
TRIVIALITY}\par
\null\vskip 1pt
\null\hskip 6cm Rolf Tarrach\par
\null\vskip 1pt
\null\hskip 3cm {\it Dep. d'ECM}, {\it Univ.
de Barcelona}\par
\null\hskip 3cm {\it Postal address: Diagonal  647}, {\it 08028
Barcelona, Spain.}\par

\null\vskip 1pt
\centerline {and {\it IFAE}}\par
\null\vskip 1cm

\centerline {\bf ABSTRACT}\par

\baselineskip=20pt

\null\hskip 1cm We critically analyze the introduction of an independent
zero momentum mode field renormalization for $\phi^4$. It leads to an
infrared divergent effective action. It does not achieve its purpose:
triviality still gives massless particles in the broken phase in the
continuum limit. It leads to an effective potential which is not the low
energy limit of the effective action.\par

\null\vskip 1cm
{\it e-mail: tarrach@ecm.ub.es}\par
\vfill\eject

\null\hskip 1cm Lattice $\phi^4$ is a theory which depends on three
parameters: two which characterize the action, $m$ and $\lambda$, and
the lattice spacing $a$. For particle physics only the
scaling region is  of interest. It is characterized by a correlation
length $\xi$ which is much larger than the lattice spacing, $\xi >>
a$. Physics, as long as $E<< a^{-1}$,
is then insensitive to the lattice spacing, and we are doing essentially
continuum physics. In the scaling region, the only one we will be
interested in, and at not too high energies, lattice $\phi^4$ is a two
parameter
field theory. It is convenient to parametrize the theory with low energy
parameters, defined in terms of renormalized Green functions at low or
zero energy. They are the renormalized mass, $m_R$, and the renormalized
coupling, $\lambda_R$. The critical theory is characterized by an
infinitive correlation length. The two parameters of the action are not
independent anymore, $m=m_c(\lambda)$. It is, at not too high energies,
a one parameter theory. Its renormalized mass vanishes, $m_R=0$.\par

\null\hskip 1cm Continuum $\phi^4$ is a two parameter theory, a priori
at least. It is defined by approaching the critical theory, and taking
at the same time the lattice spacing to zero. In doing so $a$
is traded for some external length, $L$. Defining $\hat m
(\lambda, ma)\equiv m_R a$, and recalling that for the critical theory
$\hat
m (\lambda, m_ca)=0$, one takes the continuum limit in such a way that
the continuum renormalized parameters stay finite, and in general
nonvanishing,

$$\lambda_r\equiv lim_{a \rightarrow O \atop m\rightarrow m_c}
\lambda_R(\lambda, ma)\eqno (1)$$
$$m_r\equiv lim_{a \rightarrow O \atop m\rightarrow m_c} {\hat m
(\lambda, ma) \over a}$$

This finetuned limit requires in principle three
renormalizations: field, mass and coupling. The field renormalization
takes care of the multiplicative renormalization of the Green functions.
It is normalized by the relation between Green functions and amplitudes.
The mass and coupling renormalizations are normalized by the two low
energy parameters of the theory, $m_R$ and $\lambda_R$. But, as the
theory is almost surely trivial [1], there is no interaction eventually
and the renormalized continuum coupling vanishes, $\lambda_r=0$. In
other words, the renormalized coupling $\lambda_R$
does depend in a trivial theory in an essential way on the UV cutoff $a
$ when the continuum limit is taken. There
does not exist a scaling region for interacting physics when one takes
the continuum limit. The theory is thus a one parameter theory in the
continuum limit. The parameter is, in the symmetric phase, $m_r$.\par

\null\hskip 1cm The same scenario very likely holds in the broken phase
[2]. Since the theory is now broken its renormalized vacuum expectation
value should be nonvanishing, $v_R \not= 0$. But as the theory still
only has at not too high energies two independent parameters as no new
counterterms are required, there has to be a relation between $v_R$,
$m_R$ and $\lambda_R$. It is conveniently given (up to a finite
renormalization) by

$${3 m^2_R\over {v^2_R}}= \lambda_R \eqno (2)$$

because this relation holds at tree level and because the UV divergent
parts of the counterterms satisfy it.
Eq. 2
is often taken as a definition of $\lambda_R$. In taking the continuum
limit triviality transforms it into

$${m^2_r\over {v^2_r}}=0\eqno (3)$$

As renormalized continuum parameters should be finite eq. 3 implies, as
$v_r\not=0$, that $m_r = 0$. Triviality leads to a massless particle
theory in the continuum broken phase. The steps from eq. 2 to eq. 3 have
important consequences for the high energy structure of the minimal
standard model, leading to a bound of the Higgs mass which, very
conservatively, reads $m_H< 1$TeV [3].\par

\null\hskip 1cm There have been attempts of finding a non-trivial, or,
later, a trivial massive theory in the broken phase. One of these
started from a gaussian effective potential which, after
renormalization, was non-trivial [4]. It was soon realized, however,
that one could not extend the renormalization to the effective action,
or, in other words, the effective action was UV divergent [5], and that
effective potential studies were too limited for allowing a thorough
understanding of triviality issues [6]. This is because an
unconventional field renormalization has to be checked against the UV
behaviour of the kinetic energy of the effective action, even if it
leads to a finite effective potential. These difficulties were bypassed
by introducing two field renormalizations, one for non-zero momentum
modes and one for the zero momentum mode [7]. The first one is relevant
to the effective kinetic energy, the second to the effective potential.
This conjecture soon led to the prediction of a 2 TeV Higgs [8], its
analytical finite volume study [9], the proposal of its lattice test
[10], its first lattice results [11] and a recollection of the main
ideas which lie behind a 2 TeV Higgs [12].\par

\null\hskip 1cm This attempt has produced a substantial amount of
publications on O(N) extensions, finite temperature analysis,
postgaussian corrections, perturbative approaches, renormalization group
studies, etc. to which we do not refer. It now hinges on the assumption
of performing two independent field renormalizations, which then would
lead to a trivial two parameter continuum theory, massive in the broken
phase, with $v_r$ and $m_r$ nonvanishing, contrary to the generally
accepted and very solidly founded understanding of triviality [1, 2, 3].
Then, determining $v_r$ from Fermi's constant and relating $m_r$ to
$v_r$
with a further assumption, one obtains the unconventional Higgs mass
prediction.\par

\null\hskip 1cm Now, although it is enough to check an unconventional
field renormalization for non-zero momentum modes with the effective
action, it is not enough to check an unconventional zero momentum mode
field renormalization with the effective potential, as there is one
primitively UV divergent zero momentum Green function missing from it,
because it is not proper: the connected one point function. In other
words, if there is a new renormalization it comes from new UV
divergences, and these have to show up in the one point function. In the
standard picture of symmetry breaking the one point function divergences
are not independent, but determined by the ones of the symmetric phase.
Surprisingly in none of the many publications on this subject the
unconventional zero-momentum renormalization has been checked against
the one point function to see whether it remains finite after
renormalization. It does not if the nonvanishing momentum modes are
massive. This would then lead to massless nonvanishing momentum modes,
and thus massless particles, and massive vanishing momentum modes,
unrelated to any particles. We will come to this conclusion, not by
performing a specific computation, but proving it from the general
structure of trivial quantum field theory with two field
renormalizations. Thus two independent field
renormalizations do not live up to their expectations; they are
useless and lead to pathologies.\par

\null\hskip 1cm One could of course dismiss such an undertaking offhand
on two grounds: first, one of the renormalizations, being only a zero
momentum mode renormalization, will lead to an IR divergent effective
action; and second, by renormalizing zero momentum modes differently
from
nonvanishing momentum modes the renormalized Green functions will be
discontinuous at vanishing momenta, which makes their zero momentum
values physically irrelevant. This is because the bare Green functions
cannot have a new UV divergence at zero momentum; any new zero momentum
divergence can only be IR in nature. There are two reasons why we feel
it is nevertheless worth showing that two field renormalizations are not
possible: first, the still ongoing work on the attempt started at [4];
and second, the irony of the fact that it is precisely the masslessness
of continuum, broken, trivial $\phi^4$ which allows to start the whole
attempt, as we will show.\par

\null\hskip 1cm In a QFT all the renormalizations are dictated by the UV
divergences of the Green functions, which are given by the theory
itself. In a trivial theory, because there are so many UV zeroes (all
the three and more point Green functions) and specially because in the
broken phase the zero momentum two point function vanishes in the
continuum limit, one could think of an independent renormalization of
the zero momentum mode, which would make the zero momentum Green
functions finite: it does not absorb UV divergences but cancels UV
zeroes, except for the one point function, where it absorbes UV
divergences.\par

\null\hskip 1cm In order to show how this idea is implemented, and why
it fails, let us construct the effective action in the standard way for
a $\phi^4$ field theory with two noncompeting sources, one coupled to
the zero momentum mode and the other coupled to all the other modes. No
mode should couple to both sources, as otherwise its renormalization
would be ambiguous. With euclidean metric and working in momentum space,
the generating functional is defined by

$$e^{W[{\tilde J}, j]}= N \int D {\tilde \phi}\null\hskip 2pt exp
[-S[{\tilde \phi}]+j{\tilde \phi}(0)+\int dk {\tilde
J}(-k){\tilde \phi}(k)]\eqno (4)$$

with the Lorentz invariant constraint

$${\tilde J}(0)=0\eqno (5)$$

and the normalizing factor $N$ such that $W[0,0]=0$. ${\tilde \phi}
(k)$ and ${\tilde J}(k)$ are the Fourier transformed $\phi(x)$ and
$J(x)$. An UV cutoff $a$ is in place. We assume Lorentz
invariance as we are in the scaling region and neglect scaling
violations. The thermodynamic or infinite volume limit is understood.
The theory, $\phi^4$, is symmetric, $S[{\tilde \phi}]=S[-{\tilde
\phi}]$, and $dk$ is the four-dimensional measure divided by
$(2 {\pi})^4$.
The variable source ${\tilde J}(k)$ is a smooth function which tends
to zero for large $k$, well below $a^{-1}$. The advantage of working in
momentum space is clear from eq. 5.\par

\null\hskip 1cm We are interested in the broken phase. The sign of $j$
determines which of the two equivalent $SSB$ vacua is chosen by the
theory. The standard approach starts from $W[{\tilde J}]=
lim_{j\rightarrow 0}$ $W[{\tilde J}, j]$, with $\tilde J$
unconstrained.\par

\null\hskip 1cm The effective field is given by

$${\delta W\over \delta {\tilde J}(-k)}\equiv {\tilde
\Phi}(k),\null\hskip 7pt lim_{{\tilde J}\rightarrow 0}{\tilde \Phi}(k)=
0, \null\hskip 1cm k\not= 0\eqno (6)$$

and by

$${dW\over dj}\equiv {\tilde \Phi}(0)\equiv
{\tilde \delta}(0)v, \null\hskip 7pt lim_{{\tilde
J}\rightarrow 0}v=v_j
\eqno (7)$$

where the extensive character of the ground state is made explicit by
the zero mode in form of the IR divergent volume factor ${\tilde
\delta}(0)$. Actually $v = v_j$ as $\tilde J$ only produces IR
subleading differences. SSB means that

$$lim_{j\rightarrow 0} \null\hskip 2pt v_j= v_o\not= 0 \eqno (8)$$

The effective field ${\tilde \Phi} (k)$ is a smooth function of $k$,
except at $k=0$, which vanishes for large $k$.\par

\null\hskip 1cm The effective action is given by a double Legendre
transform,

$$\Gamma [{\tilde \Phi}]\equiv W[{\tilde J}, j] - \int dk {\tilde
\Phi}(k){\tilde J}(-k)- {\tilde \Phi}(0) j \eqno (9)$$

so that $\Gamma[{\tilde \delta}(k)v_o]=0$. Also

$${\delta\Gamma\over d{\tilde \Phi}(k)}=-{\tilde J}(-k), \null\hskip
7pt lim_{{\tilde \Phi}\rightarrow 0}{\tilde J}(-k)=0,\null\hskip 1cm
k\not= 0 \eqno (10)$$

and

$${d\Gamma\over d{\tilde\Phi}(0)}=-j, \null\hskip 7pt lim_{{\tilde
\Phi}(k)\rightarrow {\tilde \delta}(k)v_o}j=0\eqno (11)$$

\null\hskip 1cm A Taylor expansion around
${\tilde\Phi}(k)={\tilde\delta}(k)v_o$ shows its character as a
generating functional of proper (one-particle irreducible, truncated,
tadpole reducible) Green functions:

$$\Gamma[{\tilde\Phi}]=\sum_{n=2}{1\over n!}{n\atop{\Pi\atop
i=1}}(\int
dk_i({\tilde\Phi}(k_i)-{\tilde\delta}(k_i)v_o)){\tilde\delta}(k_1+k_2+
\cdot\cdot k_n){\tilde\Gamma}^{(n)}(k_1, k_2\cdot\cdot k_n)\eqno (12)$$

This is the standard expression, except that
${\tilde\Phi}(k)-{\tilde\delta}(k)v_o$ is now discontinuous at $k=0$,
the discontinuity being $(v-v_o){\tilde\delta}(k)$. One can rewrite
eq. 12 as

$$\Gamma[{\tilde\Phi}]=\sum_{n=2}{1\over n!}{n\atop{\Pi\atop i=1}}(\int
dk_i({\tilde\Phi}(k_i)-{\tilde\delta}(k_i)v)){\tilde\delta}(k_1+\cdot\cdot
k_n){\tilde\Gamma}^{(n)}(k_1,\cdot\cdot k_n)$$
$$+\cdot\cdot\cdot\cdot
+{\tilde\delta}(0)\sum_{n=2}{1\over
n!}(v-v_o)^n{\tilde\Gamma}^{(n)}
(0, 0,\cdot\cdot 0)\eqno (13)$$

where only the first term, which only depends on non-zero modes, and the
last term, which only depends on zero modes, have been written out
explicitly. The zero-momentum Green functions could be IR divergent, but
this should be of no relevance to the issue at hand, which refers to the
UV structure of the theory.\par

\null\hskip 1cm Since the only $x$-independent source $J(x)$ which
satisfies eq. 5 is $J(x)=0$, the effective potential is defined
as

$$V(v_j)\equiv -{1\over{\tilde\delta}(0)}{\tilde\Gamma}[v_j
{\tilde\delta}(k)]=-\sum_{n=2}{1\over n!}(v_j -
v_o)^n{\tilde\Gamma}^{(n)}(0,\cdot\cdot\cdot 0)\eqno (14)$$

Notice that the last term of the effective action eq. 13 is IR
divergent; this is of course due to the constant source $j$. This is why
sources have to decay for large $x$, and when they are taken constant,
as when one defines the effective potential, the IR divergent volume
factor is divided out.\par

\null\hskip 1cm In fact, and as $v=v_j$, the effective action
actually contains the effective potential for a constant field. This
never happens in the standard formalism with one source, as there the
effective potential is obtained from the effective action for an
$x$-independent effective field incompatible with a source which decays
for large $x$.\par

\null\hskip 1cm Up to here the theory was bare and regularized. The bare
Green functions depend on $\tau\equiv \vert 1-{m\over m_c}\vert,
\lambda$ and $a$. They are renormalized multiplicatively

$${\tilde\Gamma}_R^{(n)}(k_1\cdot\cdot\cdot k_n)=Z^{({n\over
2})}_R \null\hskip 2pt {\tilde\Gamma}^{(n)}(k_1\cdot\cdot\cdot k_n)\eqno
(15)$$

The renormalization of the nonvanishing momentum modes thus is

$${\tilde\Phi}_R(k)=Z^{-1/2}_R{\tilde\Phi}(k), \null\hskip 1cm
k\not=0\eqno (16)$$

\null\hskip 1cm Suppose now that the theory is trivial in the known
sense [2, 13], i.e.

$$m_R\sim\tau^{1/2}\vert
ln\tau\vert^{-1/6}a^{-1}$$
$$\lambda_R\sim{32\pi^2\over 3}\vert
ln\tau\vert^{-1}\eqno (17)$$
$$Z_R\sim 1$$
$$v_R\sim C\tau^{1/2}\vert
ln\tau\vert^{1/3}a^{-1}$$

for small $\tau$,
which satisfies eq. 2 for a conveniently chosen constant $C$. Notice
that this scaling behaviour is very solidly founded, because
renormalization group improved perturbation theory is, at low energies,
and because of triviality, very reliable. Let us now take the continuum
limit in such a way that

$$a\sim \tau^{1/2}\vert ln\tau\vert^{1/12}L\eqno (18)$$

so that in the finetuned continuum limit one finds

$$m_R\sim\vert ln\tau\vert^{-1/4}L^{-1}$$
$$\lambda_R\sim {32\pi^2\over 3}\vert
ln\tau\vert^{-1}$$
$$Z_R\sim 1$$
$$v_R\sim C\vert ln\tau\vert^{1/4}L^{-1}\eqno
(19)$$

and $m_r=0$, $\lambda_r=0$ and $v_r$ diverges. The continuum limit
shown in eq. 19 can be generalized to all Green functions:

$${\tilde\Gamma}_R^{(n>2)}(k_1\cdot\cdot\cdot k_n)\sim
Z^{n/2}_A A^{(n)}(k_1\cdot\cdot\cdot k_n)$$
$${\tilde\Gamma}_R^{(2)}(k)\sim k^2+Z_A
A^{(2)}$$ $$v_R\sim Z^{-1/2}_A A\eqno (20)$$

where $A^{(n)}(k_1\cdot\cdot\cdot k_n)$ is nonvanishing, $A^{(n)}\equiv
A^{(n)}(0, 0, \cdot\cdot\cdot 0)$ and

$$Z_A\sim\vert ln\tau\vert^{-1/2}\eqno (21)$$

\null\hskip 1cm Now comes the crucial observation: in eq. 20 all the
zero-momentum proper Green functions have an UV zero, and the
connected one point function has an UV divergence in precisely such a
way that a further zero momentum field renormalization makes all of them
finite. It is given by

$${\tilde\Phi}_A(0)=Z^{-1/2}_R Z^{1/2}_A{\tilde\Phi}(0)\eqno (22)$$

This is what was (unconsciously) discovered in [4] and is the conjecture
on which the
prediction of a 2 TeV Higgs hinges. The effective action then becomes in
the continuum limit

$$\Gamma[{\tilde\Phi}_R, {\tilde\Phi}_A]={-1\over 2}\int dk
k^2{\tilde\Phi}_R(k){\tilde\Phi}_R(-k)+{\tilde\delta}(0)\sum_{n=2}
{(A-A_o)^n \over n!}A^{(n)}\eqno (23)$$

where, from eqs. 20 and 22,

$$A=lim_{a\rightarrow 0}Z^{-1/2}_R Z^{1/2}_Av$$
$$A_o=lim_{a\rightarrow 0}Z^{-1/2}_R Z^{1/2}_A v_o\eqno (24)$$

Notice that only the two terms written out in eq. 13 survive the
continuum limit. Eq. 23 is our main result. It contains a finite,
interacting effective potential, and only trivial nonvanishing momentum
physics, as put forward in references [7-12]. But it has three features
of relevance, missed in these references: First, the effective action is
still IR divergent. Sources
cannot be constant. Second, the particles of eq. 23 are still massless,
the new renormalization has not changed this. Third, the massive,
interacting effective potential of eq. 23 is not the low energy limit of
the effective action. Renormalized Green functions are discontinuous at
zero momentum, and thus is the discontinuity devoid of physics. These
results show that two independent field renormalizations do not lead to
a physically meaningful effective action, and in any case to no new
physics.
\vfill\eject

\centerline {\bf REFERENCES}\par
\null\vskip 2pt
\settabs \+1aaaaa&\cr

\+1.&K.G. Wilson and J. Kogut, Phys. Rep. \underbar {C12} (1974)
75\cr\par
\+&M. Aizenman, Phys. Rev. Lett. \underbar {47} (1981) 1\cr\par
\+&J. Fr\" ohlich, Nucl. Phys. \underbar {B200} [FS4] (1982) 281\cr\par
\+&A. Sokal, Ann. Inst. H. Poincar\'e \underbar {37} (1982) 317\cr\par
\+&D.J. Callaway and R. Petronzio, Nucl. Phys. \underbar {B240} (1984)
577\cr\par
\+&C.B. Lang, Nucl. Phys. \underbar {B265} (1986) 630\cr\par
\+&M. L\" uscher and P. Weisz, Nucl. Phys. \underbar {B290} (1987)
25\cr\par
\+2.&M. L\" uscher and P. Weisz, Nucl. Phys. \underbar {B295} (1988)
65\cr\par
\+3.&R. Dashen and H. Neuberger, Phys. Rev. Lett. \underbar {50} (1983)
1897\cr\par
\+&J. Kuti, L. Lin and Y. Shen, Phys. Rev. Lett. \underbar {61} (1988)
678\cr\par
\+4.&M. Consoli and A. Ciancitto, Nucl. Phys. \underbar {B254} (1985)
653\cr\par
\+&P.M. Stevenson and R. Tarrach, Phys. Lett. \underbar {B176} (1986)
436\cr\par
\+5.&S. Paban and R. Tarrach, Phys. Lett. \underbar {B213} (1988)
48\cr\par
\+&J. Soto, Nucl. Phys. \underbar {B316} (1989) 141\cr\par
\+&B. Rosenstein and A. Kovner, Phys. Rev. \underbar {D40} (1989)
504\cr\par
\+6.&R. Tarrach, Phys. Lett. \underbar {B262} (1991) 294\cr\par
\+7.&V. Branchina, P. Castorina, M. Consoli and D. Zappala, Phys. Lett.
\underbar {B274} (1992) 404\cr\par
\+&V. Branchina, M. Consoli and N.M. Stivala, Zeitsch. Phys. \underbar
{C57} (1993) 251\cr\par
\+8.&R. Iba\~nez-Meier and P.M. Stevenson, Phys. Lett. \underbar {B297}
(1992) 144\cr\par
\+&M. Consoli, Phys. Lett. \underbar {B305} (1993) 78\cr\par
\+9.&U. Ritschel, Phys. Lett. \underbar {B318} (1993) 617\cr\par
\+10.&M. Consoli and P.M. Stevenson, Zeitsch. Phys. \underbar {C63}
(1994) 427\cr\par
\+11.&A. Agodi, G. Andronico and M. Consoli, Zeitsch. Phys. \underbar
{C66} (1995) 439\cr\par
\+12.&M. Consoli and P. Stevenson, ``Resolution of the $\lambda\phi^4$
puzzle and a 2 TeV Higgs boson''\cr\par
\+&Rice University preprint,
DE-FG05-92ER40717-5, July 1993\cr\par
\+13.&E. Br\'ezin, J.C. Le Guillou and J. Zinn-Justin, ``Field
theoretical approach to critical\cr\par
\+&phenomena'' in ``Phase transitions and
critical phenomena'', vol. 6, eds. C. Domb,\cr\par
\+&M.S. Green (Academic Press,
London, 1976)\cr\par
\vfill
\eject

ACKNOWLEDGEMENTS\par

\null\hskip 1cm I thank Aneesh Manohar for hospitality at the Dept. of
Physics, UCSD, where this work was done. Julius Kuti's comments have
been essential in streamlining my understanding of triviality. The
endless discussions with Paul Stevenson have helped me in sharpening my
criticism. I thank him very much for them. I am only sad of not yet
having
been able to convince him that there is only one Z. Josep Taron and
Domenec Espriu have made the manuscript less criptic.\par

\null\hskip 1cm I acknowledge financial support under CICYT contract
AEN95-0590, DGICYT contract PR95-015 and CIRIT contract GRQ93-1047.\par
\vfill
\eject

\end